\documentclass[aps, pre, reprint, nobibnotes, amsmath, amssymb, aps]{revtex4-2}

\usepackage{graphicx}
\usepackage{dcolumn}
\usepackage{bm}

\newcommand{\Eprint}[1]{}

\makeatletter
\newcommand{\href@noop}{}
\makeatother

\usepackage[svgnames]{xcolor}
\usepackage{bbold}

\RequirePackage[breaklinks]{hyperref}
\hypersetup{colorlinks=true, urlcolor=MidnightBlue,
	linkcolor=MidnightBlue, citecolor=MidnightBlue}

\newcommand{\arxiv}[1]{{arXiv}:\href{http://arxiv.org/abs/#1}{#1}}

\DeclareMathOperator{\diag}{diag}

\usepackage{tikz}
\usetikzlibrary{decorations.pathmorphing}

\protected\def\[#1\]{\begin{align}#1\end{align}}

\newcommand{\Left}{\mathopen{}\mathclose\bgroup\left}
\newcommand{\Right}{\aftergroup\egroup\right}

\newcommand{\sfrac}[2]{\ensuremath{\mathchoice{\tfrac{#1}{#2}}{\tfrac{#1}{#2}}{\frac{#1}{#2}}{\frac{#1}{#2}}}}
\newcommand{\half}{\sfrac{1}{2}}

\newcommand{\isodate}[3]{%
	\number#1-%
	\ifcase#2\or
	  01\or 02\or 03\or 04\or 05\or 06\or
	  07\or 08\or 09\or 10\or 11\or 12\fi -%
	\ifcase#3\or
	  01\or 02\or 03\or 04\or 05\or 06\or 07\or 08\or 09\or 10\or
	  11\or 12\or 13\or 14\or 15\or 16\or 17\or 18\or 19\or 20\or
	  21\or 22\or 23\or 24\or 25\or 26\or 27\or 28\or 29\or 30\or
	  31\fi}
\newcommand{\isotoday}{\isodate{\year}{\month}{\day}}

\newcommand{\A}{{\text{A}}}
\newcommand{\B}{{\text{B}}}
\newcommand{\C}{{\text{C}}}
\newcommand{\D}{{\text{D}}}
\newcommand{\ex}{{\text{ex}}}
\newcommand{\eq}{{\text{eq}}}
\newcommand{\rev}{{\text{rev}}}
\newcommand{\inst}{{\text{inst}}}
\newcommand{\p}{{p}}

\begin{document}

\title{Dual thermodynamic ensembles, relative entropies, and excess free energy}

\author{Gavin E.\ Crooks}
\affiliation{B.I.T.S. (Berkeley Institute for Science and Technology)}
\email{gec@threeplusone.com}
\date{\isotoday}

\begin{abstract}
  It has long been known that the relative entropy of a non-equilibrium ensemble to the corresponding equilibrium ensemble is the excess free energy.  We show that the reverse relative entropy also has a
 thermodynamic interpretation: it is the excess free energy of a dual ensemble in which the roles of energy and entropy are interchanged.
\end{abstract}

\maketitle

We consider an ensemble $\B$ of a system thermally coupled to an idealized heat bath at constant inverse temperature $\beta=1/k_\text{B} T$, with energy spectrum $E_\B(x)$, and probabilities $\p_\B(x)$. This ensemble is out of thermal equilibrium, and therefore the probabilities are not given by the canonical ensemble of equilibrium statistical mechanics~\cite{Gibbs1902a},
\begin{align}
\p_\B(x) \neq \frac{e^{- \beta E_\B(x)} }{Z_\B} \ .
\end{align}

We define the ensemble $\A$ as the equilibrium thermodynamic ensemble of the same system, coupled to the same heat bath, with the same energy spectrum $E_\A(x) = E_\B(x)$, but with canonical probabilities,
\begin{align}
\p_\A(x) = \frac{e^{- \beta E_\A(x)} }{Z_\A}
= e^{-\beta E_\A(x) + \beta F_\A} \ .
\end{align}
Here $Z_\A= \sum_x \exp(-\beta E_\A(x))$ is the partition function, and $F_\A$ is the free energy of ensemble $\A$~\cite{Helmholtz1882a}, defined as $\beta F_\A = \beta \langle E_\A\rangle_\A - S_\A$, where $S$ is the entropy of the ensemble in nats~\cite{Clausius1865a,Boltzmann1872a,Shannon1948a,Jaynes1957a}, $S_\A = -\sum p_\A(x) \ln p_\A(x)$, and $\langle E_\A\rangle_\A$ is the average energy of ensemble~$\A$.
\[
\langle E_\A\rangle_\A & = \sum_x \p_\A(x) E_\A(x)
\]

For a canonical ensemble the partition function and free energy are related by $\beta F_\A = -\ln Z_\A$, but for a non-equilibrium non-canonical ensemble, this relation no longer holds,
$\beta F_\B \neq -\ln Z_\B$.

The relative entropy (KL-divergence)~\cite{Kullback1951a,Cover2006a} of the ensemble $\B$ relative to the equilibrated ensemble $\A$ is the difference in free energy between them, $\beta F_\B - \beta F_\A$, where $ F_\B$ is the non-equilibrium free energy of ensemble $\B$~\cite{Bernstein1972a, Shaw1984a, Gaveau1997a, Qian2001a, Hatano2001a, Vaikuntanathan2009a, Hasegawa2010a, Takara2010a, Esposito2011a, Crooks2011a, Sivak2012a, Deffner2012a}. (For a summary of the history of this result, see~\cite{Sivak2012a}.)
\begin{align}
D(\B\|\A) &= \sum_x \p_\B(x) \ln \frac{\p_\B(x)}{\p_\A(x)}
\label{eqD} \\
& = \sum_x \p_\B(x) \ln \p_\B(x)  - \sum_x \p_\B(x) \ln \p_\A(x) \notag\\
& = \sum_x \p_\B(x) \ln \p_\B(x)  \notag
\\
& \qquad -\sum_x \p_\B(x) (-\beta E_\A(x) +\beta F_\A)
\notag \\
& = -S_\B + \langle \beta E_\B \rangle_\B  - \beta F_\A
\notag\\
& = \beta F_\B - \beta F_\A \notag\\
&= \beta F^{\ex}_\B
\notag
\end{align}
Here we define the free energy of a non-equilibrium ensemble analogously to that of an equilibrium ensemble: $\beta F_\B = \beta \langle E_\B\rangle_\B - S_\B$.
Note that relative entropy is non-negative, being zero only if the ensembles are identical. Thus the excess free energy (the difference between non-equilibrium and corresponding equilibrium values) is positive and the free energy is minimized in thermodynamic equilibrium, as expected.

To justify this definition of non-equilibrium free energy, we construct a reversible thermodynamic path connecting $\B$ to $\A$~\cite{Hasegawa2010a, Takara2010a}.
We introduce a third ensemble, $\C$, that has the same probabilities as $\B$, $\p_\B(x) = \p_\C(x)$, but whose energy spectrum is chosen so that $\C$ is in thermodynamic equilibrium:
\begin{align}
   \beta E_\C(x) = - \ln \p_\B(x) + \varepsilon_c
\end{align}
Here $\varepsilon_c$ is an arbitrary energy offset that can be absorbed into the normalization constant. It does not affect the probability distribution of $\C$, but it does affect what we think the energy and free energy are. Since the maximum probability of any state is unity, it follows that $\varepsilon_c$ is the lowest energy the ground state could have. We will return to this offset later, Eq.~\eqref{offset}.

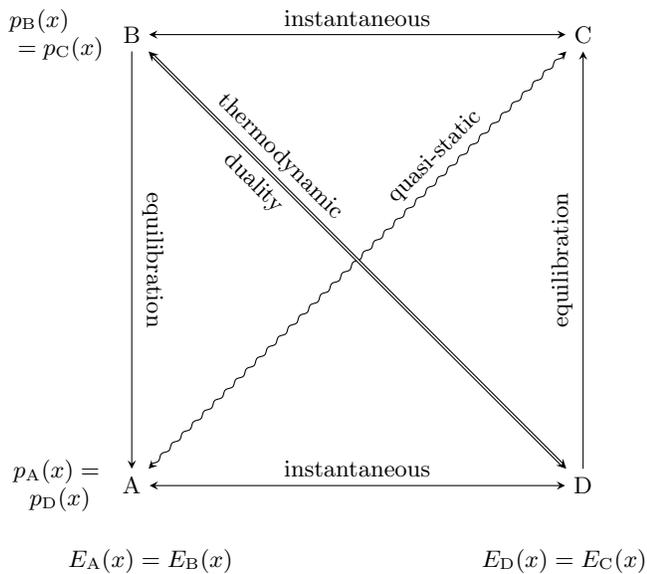
\begin{figure}
\begin{center}
\begin{tikzpicture}[>=stealth]
  \node () at (-1,0) [align=left] {$p_\A(x)= $ \\ $\ \  p_\D(x)$};
  \node () at (-1,6) [align=left] {$p_\B(x)$\\ $\ = p_\C(x)$};
  \node () at (0.25,-1) {$E_\A(x) = E_\B(x)$};
  \node () at (5.75,-1) {$E_\D(x) = E_\C(x)$};

  \node (A) at (0,0) {A};
  \node (B) at (0,6) {B};
  \node (C) at (6,6) {C};
  \node (D) at (6,0) {D};

  \draw[->] (B) -- (A)
    node[midway, xshift=+8pt, rotate=270, anchor=center]{equilibration};

  \draw[->] (D) -- (C)
      node[midway, xshift=-8pt, rotate=90, anchor=center]{equilibration};

  \draw[<->] (B) -- (C)
    node[midway, above]{instantaneous};

  \draw[<->] (A) -- (D)
        node[midway, above]{instantaneous};

  \draw[<->, decorate, decoration={snake, amplitude=0.2mm, segment length=2mm}] (A) -- (C)
       node[midway, xshift=-10pt, rotate=45, anchor=center]{\hspace{12em}quasi-static};

  \draw[<->, double] (D)  -- (B)
       node[midway, xshift=10pt, rotate=-45, anchor=center]{thermodynamic\hspace{12em}  } node[midway, xshift=-10pt, rotate=-45, anchor=center]{duality\hspace{9em}  };

\end{tikzpicture}
\end{center}
\caption{Thermodynamic duality of non-canonical ensembles. Ensemble $\B$ is out-of-thermodynamic equilibrium and non-canonical, whereas $\A$ is the equilibrium ensemble the system would relax to if left undisturbed. Ensembles $\B$,$\C$ and $\A$,$\D$ share the same probability distributions respectively, while $\A$,~$\B$ and $\C$,~$\D$ share the same energy spectrum. Consequently, whereas the relative entropy $D(\B\|\A)$ is the excess free energy of ensemble $\B$, the reverse relative entropy $D(\A\|\B)= D(\D\|\C)$ is the excess free energy of the thermodynamically dual ensemble $\D$.
}
\end{figure}

The path from $\B$ to $\A$ proceeds in two steps. First, we instantaneously change the energy levels from $E_\B$ to $E_\C$, stabilizing the non-equilibrium distribution~\cite{Hasegawa2010a}. Since the probabilities are unchanged, there is no entropy change, and the work is purely energetic:
\begin{align}
   \beta \langle W^\text{inst}_{\B\rightarrow\C} \rangle &= \sum_x \p_\B(x) \bigl(\beta E_\C(x) - \beta E_\B(x)\bigr) \notag \\ \notag & = \langle \beta E_\C\rangle_\C - \langle  \beta E_\B\rangle_\B  \\&= \beta F_\C - \beta F_\B
\end{align}
This instantaneous transformation is thermodynamically reversible in the sense that there is no change in entropy, provided that we do not allow any time for the system to relax and dissipate energy.

Second, we convert the equilibrium ensemble $\C$ to $\A$ by a quasi-static transformation. The work equals the difference in free energies:
\begin{align}
    \beta W^\text{rev}_{\C\rightarrow \A} = \beta F_\A - \beta F_\C
\end{align}
The total reversible work along this path is
\[
\beta \langle W_{\B\rightarrow\A}^\rev \rangle &= (\beta F_\C - \beta F_\B) + (\beta F_\A - \beta F_\C) \\
&= \beta F_\A - \beta F_\B = -\beta F^{\ex}_\B \ ,
\]
confirming that the excess free energy is the work extractable by a reversible process connecting the non-equilibrium and equilibrium ensembles.

An independent justification follows from the Jarzynski identity~\cite{Jarzynski1997a,Jarzynski1997b,Crooks1998a},
\[
\Left\langle e^{-\beta W} \Right\rangle_{\A, \Lambda} & = e^{-\beta \Delta F_{\Lambda}} \ ,
\]
where $W$ is the work done on a system initially in equilibrium $\A$, and $\Delta F_\Lambda$ is the change in equilibrium free energy induced by the protocol $\Lambda$. By Jensen's inequality the average excess work is non-negative,
\[
\Left\langle \beta W  \Right\rangle_{\A,\Lambda}  \geq  \beta \Delta F_\Lambda \ ,
\]
which is essentially a statement of the second law.
We can extend this to non-equilibrium initial conditions by writing the Jarzynski identity explicitly,
\[
 \sum_{x_0, X} p_\A(x_0)\ p(X|\Lambda, x_0)\ e^{-\beta W[X,\Lambda] } & = e^{- \beta \Delta F_\Lambda} \ ,
\]
where $x_0$ is the initial state and $X$ is the trajectory during the driving process, and then substituting a different initial distribution $p_\B$,
\[
\sum_{x_0, X} p_\B(x_0) p(X|\Lambda,x_0)\ e^{-\beta W - \ln \frac{p_\B(x_0)}{p_\A(x_0) }  }  & = e^{- \beta \Delta F_\Lambda} \ .
\]
This yields a Jarzynski-like identity valid for any initial distribution,
\[
\Left\langle e^{-\beta W  - \ln \frac{p_\B(x_0)}{p_\A(x_0) }  } \Right\rangle_{\B, \Lambda} & = e^{- \beta \Delta F_\Lambda} \ ,
\]
from which Jensen's inequality gives
\[
\Left\langle \beta W \Right\rangle_{\B, \Lambda} & \geq \beta \Delta F_\Lambda - \beta F^{\ex}_\B \ .
\]
If the protocol $\Lambda$ leaves the equilibrium ensemble unchanged ($\Delta F_\Lambda = 0$), then starting from a non-equilibrium initial condition the average work is bounded by the excess free energy:
\[
\Left\langle \beta W \Right\rangle_{\B, \Lambda} & \geq  - \beta F^{\ex}_\B = \beta F_\A - \beta F_\B \ .
\]
(This derivation is a minor modification of the discussion found in~\cite{Vaikuntanathan2009a}.)

These preceding arguments show that it is proper to define the free energy of a non-equilibrium ensemble as $\beta F =\beta \langle E \rangle - S$, and that this free energy is related to the relative entropy between non-equilibrium and corresponding equilibrium ensembles~\eqref{eqD}.
Howsoever, we are intending to give a thermodynamic interpretation for the reverse, or dual, relative entropy $D(\A\|\B)$. To this end we introduce a fourth ensemble $\D$. This ensemble has the same probability distribution as $\A$, and the same energy spectrum as $\C$.
\begin{align}
\p_\D(x) & = \p_\A(x) = \frac{ e^{- \beta E_\A(x)} }{Z_\A}
\\
\beta E_\D(x) & = \beta E_\C(x) = - \ln \p_\B(x) + \varepsilon_c
\end{align}
This new ensemble $\D$ is thermodynamically dual to ensemble $\B$ in that the roles of entropy and energy have been interchanged: the probabilities of $\D$ are defined by the energies of $\B$, and the energies of $\D$ are defined by the probabilities of $\B$.
The interrelation of these four ensembles is summarized by the diagram in Fig.~1.

We are now in a position to give a thermodynamic interpretation of $D(\A\|\B)$. Since the probability distribution of $\A$ is the same as $\D$, and that of $\B$ the same as $\C$, it follows that $D(\A\|\B) = D(\D\|\C)$. But since $\C$ is the canonical equilibrium that the non-canonical ensemble $\D$ would relax to (they have the same energy spectrum and bath temperature), it follows that $D(\D\|\C)$ is the excess free energy of the dual ensemble $\D$.
\begin{align}
D(\A\|\B) = D(\D\|\C) = \beta F_\D - \beta F_\C = \beta F^\text{\ex}_\D
\end{align}

To make the concept of a thermodynamically dual ensemble more concrete, we can introduce the notation $\B^\star=\D$. The corresponding equilibrated ensemble can be denoted as $\B^\eq = \A$, and the instantaneously stabilized ensemble as $\B^\inst = \B^\star{}^\eq = \C$.
\begin{align}
\p_{\B^\star}(x) &= \frac{ e^{- \beta E_\B(x)} }{Z_\B}
\\
\beta E_{\B^\star}(x) &= - \ln \p_\B(x) + \beta F_{\B^\eq}
\label{offset}
\end{align}
We have also fixed the energy offset $\varepsilon_{\B^\star}=\beta F_{\B^\eq}$. This is convenient (but not thermodynamically necessary) as it ensures that the free energies of the equilibrated and stabilized ensembles are the same, $\beta F_{\B^\inst}= \beta F_{\B^\eq}$, and therefore that the thermodynamic duality is an involution $\B^{\star\star} = \B$. By this definition, ensembles in thermodynamic equilibrium are self-dual $\B^{\eq{}\star}=\B^\eq$.

Non-equilibrium ensembles are inherently dynamical, characterized by persistent probability currents and entropy production. It is therefore natural to ask how thermodynamic duality extends to the underlying stochastic dynamics.
We will work within the framework of continuous-time Markov chains, the essentials of which we will now summarize~\cite{Norris1997a}.

Let $R$ be the irreducible transition rate matrix of a continuous-time Markov chain. All of the off-diagonal elements are non-negative and columns sum to zero (due to conservation of probability).
\[
R p &= \dot{p} \\
R_{ij} &\geq 0 \text{ for } i\neq j \\
\sum_i R_{ij} &= 0 \text{ for all } j
\]
The last condition can also be written as $\boldsymbol{1}^\top\, R = \boldsymbol{0}^\top $.

The stationary distribution $\pi$ of $R$ is invariant in time, $R \pi = \boldsymbol{0}$. The time reversal $\widetilde{R}$ of the rate matrix $R$ is~\cite{Kolmogorov1936a,Kelly1979a,Norris1997a}
\[
\widetilde{R} = \diag(\pi)\, R^\top\, \diag(\pi)^{-1} \ .
\]
A rate matrix that is symmetric under time reversal $R=\widetilde{R}$ is said to be detailed balanced~\cite{Dirac1924a,Kolmogorov1937a}, and the resulting dynamics is microscopically reversible~\cite{Tolman1924a}.

Given a rate matrix, we can also define a conservative probability flow matrix $\Phi$~\cite{Seabrook2023a,Sawchuk2026a},
\[
 \Phi & = R  \diag(\pi) \ .
\]
The off-diagonal entries give the stationary flow of probability mass from one state to another. All non-diagonal elements are non-negative and all rows and columns sum to zero:
\[
\Phi_{ij} \geq 0 \text{ for } i\neq j  \ , \quad
\boldsymbol{1}^\top\, \Phi = \boldsymbol{0}^\top \ ,
\quad
\Phi\,  \boldsymbol{1}= \boldsymbol{0} \, .
\]
Up to normalization, the flow matrix is thus a transition rate matrix with a uniform stationary state, and can be seen as the continuous-time equivalent of a discrete-time doubly stochastic matrix.
Consequently, the time reversal of the flow matrix is given by the transpose of the matrix, $\widetilde{\Phi} = \Phi^\top$.

The flow matrix can be decomposed into the sum of time-symmetric and skew-symmetric components.
\[
\Phi = \underbrace{\tfrac{1}{2}(\Phi+\Phi^\top)}_{\Phi_{\text{sym}}}
\;+\;
\underbrace{\tfrac{1}{2}(\Phi-\Phi^\top)}_{\Phi_{\text{skew}}}
\]
In stochastic thermodynamics, $2\Phi_{\text{sym}}$ is known as the traffic, and $2\Phi_{\text{skew}}$ the probability current~\cite{Maes2020a}.

The entropy production rate $\sigma$ can also be expressed in terms of the flow matrix alone~\cite{Schnakenberg1976a},
\[
 \sigma
  & =\sum_{i\ne j}p_jR_{ij}\log\frac{p_jR_{ij}}{p_iR_{ji}}
  =\sum_{i\ne j}\Phi_{ij}\log\frac{\Phi_{ij}}{\Phi_{ji}}
  \ .
  \label{entprod}
\]

Suppose we are given the rate matrix $R_\B$ representing the steady-state dynamics of ensemble $\B$, with corresponding flow matrix $\Phi_{\B} = R_\B \diag(p_\B)$. Since the entropy production rate~\eqref{entprod} depends only on the flow matrix, a natural requirement is that the dual ensemble $\D$ share the same flow matrix as $\B$, $\Phi_{\D} = \Phi_{\B}$, and therefore the same entropy production rate. Given this constraint and a new positive stationary distribution $p_\D$, the rate matrix of $\D$ is uniquely determined:
\[
R_\D &= \Phi_{\D}  \diag(p_\D)^{-1} \   \\
 & \quad = R_\B \diag(p_\B) \diag(p_\D)^{-1}  \notag
    \ .
    \label{transform}
\]
This new matrix has non-negative off-diagonal elements, conserves probability,
\[
\boldsymbol{1}^\top R_\D &= \boldsymbol{1}^\top \Phi \diag(p_\D)^{-1} = \boldsymbol{0}^\top \, ,
\]
and has the stationary distribution $p_\D$,
\[
R_\D\, p_\D &= \Phi \diag(p_\D)^{-1} p_\D =  \Phi \boldsymbol{1} = \boldsymbol{0} \ .
\]

A continuous-time Markov chain can also be characterized by the holding times, the mean time the system dwells in each state before making a transition to another state, and the jump probabilities, the probability that the system transitions from state $j$ to state $i$, given that a transition has occurred.
The dynamical transform~\eqref{transform} preserves the jump probabilities $\Pi_{ij} = -\Phi_{ij}/\Phi_{jj}$ and only dilates the holding times $\tau_j = -p_j/\Phi_{jj}$ to obtain the desired stationary distribution.

The instantaneously stabilized ensembles $\A$ and $\C$ are in thermodynamic equilibrium, the defining feature of which is that we cannot distinguish a direction of time, since entropy is neither increasing nor decreasing. The dynamics must be statistically time reversal invariant, and therefore the flow matrices symmetric. The nearest symmetric flow matrix to $\Phi_\B$ in the Frobenius norm is the time symmetrization,
$\Phi_\A = \Phi_\C = \half(\Phi_\B + \Phi^\top_\B)$,
since the symmetric part of a flow matrix is itself a flow matrix.

We can now express $R_\A$ and $R_\C$ in terms of the rate matrix $R_\B$.
\[
R_\A &= \half(R_\D + \widetilde{R}_\D)
\notag \\
 & = \half(R_\B \diag(p_\B) \diag(p_\D)^{-1}
\notag \\ &  \qquad +
\diag(p_\B) R_\B^\top \diag(p_\D)^{-1}
)
\]
\[
R_\C &= \half(R_\B + \widetilde{R}_\B) \notag
\\ &
 = \half(R_\B + \diag(p_\B) R^\top_\B \diag(p_\B)^{-1})
\]

Note that since $R_\A$ and $R_\C$ share a flow matrix, they are interrelated by the same dynamical time-dilation transform as are $R_\D$ and $R_\B$.

The thermodynamic duality between forward and reverse relative entropy also manifests in the theory of hyperensembles, probability distributions over probability distributions~\cite{Crooks2007b, Dixit2013a, Naudts2007a, Hummer2015a, Naudts2011a, Dixit2015a, Gao2016a}. When constructing a hyperensemble from limited information, the choice of divergence determines the form of the resulting distribution over states. Minimizing the forward relative entropy (maximum entropy principle) yields an entropic prior over distributions, whereas minimizing the reverse relative entropy yields a Dirichlet distribution. Essentially, the entropic and Dirichlet hyperensembles are themselves duals under the interchange of energy and entropy.

In variational inference one typically minimizes the reverse relative entropy $D(q\|p)$~\cite{Blei2017a,Kingma2014a}, whose well-known mode-seeking behavior contrasts with the mean-seeking behavior of the forward divergence.
Thermodynamic duality provides a physical interpretation of this asymmetry: the forward and reverse relative entropies are excess free energies of dual ensembles in which the roles of energy and entropy are interchanged.

Herein we have shown how to construct dual thermodynamic ensembles where the roles of entropy and energy are interchanged. These dual ensembles have interesting connections to inference and machine learning and show that both the forward and reverse relative entropies of an ensemble relative to its equilibrium are excess free energies.
These thermodynamically dual ensembles can, in principle, be realized experimentally.
It remains to be seen if they can be realized in practice.

Acknowledgments: I thank David Limmer and 
Jordan Sawchuk for incisive discussions and helpful suggestions.

{
\hbadness=10000 % Suppress hbox underfull warnings in bibliography.
\bibliography{gec_bibliography}
}

\end{document}